\documentstyle[12pt]{article}
\begin{document}
\begin{titlepage}
\title{Long-range interactions \\ in a one-dimensional electron system $^{*}$}
\author{Stefano Bellucci \\
           \\{\small \it Laboratori Nazionali di Frascati, INFN
P.O. Box 13, I-00044 Frascati, Roma (Italy)}}
\date{}% \today{$\mbox{}$ \vspace*{0.3truecm} \\} 
\maketitle
\thispagestyle{empty}      \vspace*{1truecm}
\begin{abstract}
We study the unscreened Coulomb interaction in a one-dimensional
electron system at low-energy. We use renormalization group methods
and a GW approximation, in order to analyze the model. This yields
both a strong wavefunction renormalization and
a renormalization of the Fermi velocity. The significance of the
effects depends on the filling level of the Fermi system. Despite
the long-range character of the interaction, the system still
falls into the Luttinger liquid universality class, since the
effective couplings remain bounded at arbitrarily low energies.
\end{abstract} 
                    \vspace*{3truecm}
* To be published in the Proceedings of
the 6th International Conference on
Path-Integrals from peV to TeV,
Florence, Italy, 25-29 August 1998
\vfill
\begin{center}
October 1998
\end{center}
\end{titlepage}  

The motivation for our interest in the study of
one-dimensional electron systems is the
recent experimental availability of single fullerene
nanotubes \cite{nano}. In one spatial dimension the
Luttinger liquid concept replaces the Fermi liquid picture.
This is usually treated assuming a local (short-range)
interaction \cite{sol,hal,vt,gz}.
For isolated fullerene tubules the bare interaction is not screened.
The effects that the unscreened Coulomb
interaction (CI) have on the properties of the one-dimensional
system have been assessed recently \cite{bs}.

In the bosonization approach for the Luttinger liquid behavior
the CI is singular at small momentum
transfer, hence a screening of the Coulomb potential by external
charges is required \cite{fisher}.
We use renormalization group (RG)
methods to find the low-energy effective theory of the $1/|x|$ 
interaction. The main
point that we want to address is the stability of the CI
in the RG framework, and whether it falls into
the Luttinger liquid universality class.
We introduce the dynamical screening due to plasmons, through a GW
approximation \cite{gw}.
The same approach has been also tested in the study of the crossover 
from Fermi liquid to Luttinger liquid behavior \cite{dicastro},
as well as in the study of singular interactions in dimension $1
< d \leq 2$ \cite{wen}.

We consider a one-dimensional one-band model with
an interaction hamiltonian
\begin{equation}
H_{int} = \frac{e^2}{8\pi } \int dx dx'\; \Psi^{+} (x) \Psi (x)
\frac{1}{|x - x'|} \Psi^{+} (x') \Psi (x')\; ,\;
\label{ham}
\end{equation}
where $\Psi (x)$ is the electron annihilation field operator.
The problem of having a nonlocal operator in (\ref{ham}) (RG methods
usually deal with local operators),
can be circumvented by introducing a local
auxiliary field $\phi (x)$ that propagates the CI \cite{bs}.

Focusing on the scaling behavior of the irreducible
functions as the bandwidth cutoff $E_c$ is sent towards the
Fermi level, $E_c \rightarrow 0$, yields the 1-loop electron
self-energy \cite{bs}
\begin{eqnarray}
i \Sigma (k,0)   \approx  i \frac{e^2}{4\pi^2} k \; \log E_c
\; .\;
\label{ren}
\end{eqnarray}
However, quantum corrections strongly modify the propagator of
the $\phi (x)$ field.
Taking into account the particle-hole processes contributing to
the 1-loop self-energy of $\phi (x)$, leads to\footnote{The ultraviolet
cutoff $\Lambda $ for excitations along the $y$ and $z$
transverse directions is needed, when projecting the
three-dimensional interaction down to the one-dimensional
system.}
\begin{equation}
i \langle \phi (k,\omega ) \; \phi (-k,-\omega ) \rangle =
 1 / \left( -\frac{2 \pi}{\log(|k|/\Lambda) } + \frac{e^2}{\pi}
\frac{v_F k^2} {v_F^2 k^2 - \omega_k^2 } \right)
\label{prop2} \; .\;
\end{equation} 

We use the scalar propagator
(\ref{prop2}) in the renormalization of the Fermi velocity and
the electron wavefunction.
This corresponds to taking the leading order in a $1/N$
expansion, in a model with $N$ different electron flavors.  In
our case, such an approximation to the self-energy is also
justified since it takes into account, at each level in
perturbation theory, the most singular contribution at small
momentum transfer of the interaction.

This approximation yields for the renormalized electron propagator
\cite{bs}
\begin{eqnarray}
G^{-1}(k,\omega_k) & \approx &
Z^{-1}_{\Psi} \; (\omega_k - v_F  k) -
Z^{-1}_{\Psi} \; (\omega_k - v_F k)  \int^{E_c} 
 \frac{dp}{|p|} \left( 1 - \frac{1+f(p)}{2 \sqrt{f(p)}}
\right)                  \nonumber               \\ &  &  -
Z^{-1}_{\Psi} \; k \; \frac{e^2}{4\pi^2 }\int^{E_c} 
 \frac{dp}{|p|} 
   \frac{\sqrt{f(p)} - 4/3  +  1/ \left( 3  f(p)^{3/2} \right)  }
  { \left( 1-f(p) \right)^2  } \; ,\;
\label{gren}
\end{eqnarray}
where $f(p) \equiv 1 - e^2 \log(|p|/\Lambda) /(2 \pi^2 v_F) \;  $ and
$Z^{1/2}_{\Psi}$ represents the scale of the
bare electron field compared to that of the cutoff-independent
electron field
\begin{equation}
\Psi_{bare}(E_c) = Z^{1/2}_{\Psi} \Psi \; .\;
\end{equation}

The RG flow equations for the cutoff-dependent
effective parameters, corresponding to the behavior of the quantum
theory as $E_c \rightarrow 0$, can be derived from (\ref{gren})
\begin{eqnarray}
E_c \frac{d}{dE_c }\:  \log \: Z_{\Psi}(E_c)  & = & 
   \frac{1+f(E_c)}{2 \sqrt{f(E_c)}} - 1      \label{zflow}   \\
E_c \frac{d}{dE_c } \: v_F (E_c)  & = &  - \frac{e^2}{4\pi^2}
    \frac{\sqrt{f(E_c)}  -  4/3  +  1/ \left(3 f(E_c)^{3/2} \right)}
    {\left( 1-f(E_c) \right)^2} \; . \;
\label{vflow}
\end{eqnarray}
One can check that Eq.
(\ref{vflow}) leads to an enhancement of the Fermi velocity at
low energies, which goes in the direction of screening the
effective interaction. Eq. (\ref{zflow}) also corresponds to a
sensible effect, as the leading behavior is that of suppressing
the electron quasiparticle weight. The fast wavefunction
renormalization we found \cite{bs}, seems to imply a non-algebraic
behavior of the electron propagator with regard to the frequency
and momentum dependence, what is reminiscent of similar features
found for some correlation functions in the bosonization
approach \cite{schulz2}.

At the level of the $1/N$ approximation we consider, the three-point
vertex only gets the cutoff dependence given by the wavefunction
renormalization in (\ref{zflow}). This means that the electron
charge is not renormalized at low energies in our local field
theory framework. The behavior of the effective interaction is,
therefore, completely encoded in Eq. (\ref{vflow}).

The RG equation for the effective coupling constant $g \equiv
e^2 /(4\pi^2 v_F)$ is
\begin{equation}
E_c \frac{d}{dE_c } g(E_c) = \frac{1}{4(\log \: E_c)^2} \left(
 \sqrt{f(E_c)} - \frac{4}{3} + \frac{1}{3f(E_c)^{3/2}} \right) \; .\;
\label{gflow}
\end{equation}
This equation actually controls all the physical properties of
the electron system. In our model the effective
coupling constant $g$ displays marginal behavior. However, the
logarithmic corrections to scaling 
in (\ref{gflow}) reduce the flow in the
infrared. The flow of $g (E_c)$ 
given by (\ref{gflow})
is arrested close to some fixed-point value, after
which it becomes insensitive to further scaling in the infrared.

We conclude that the CI remains long-ranged
in the low-energy effective theory, while the effective coupling
has a stable flow in the infrared. This means that the
one-dimensional system falls into the Luttinger liquid
universality class.
This does not support any phase transition concerning the strength of the
interaction. However, a reduction of the
effective couplings, depending on the band structure,
could be relevant for the phenomenology of
chiral one-dimensional electron systems and their application to
fullerene nanotubes \cite{fisher}.

Notice that a significant renormalization of $v_F$
may be present in small chains, contributing to explain
the insulator-metal transition by the effect of the Coulomb
interaction observed in the exact diagonalization of finite
rings \cite{poil}. The study of small finite-size systems \cite{last}
also indicates a reduction of the electron correlations similar to
our findings in the RG framework.
A generalization of our research, to include the case of different
values of the running parameter $v_F$ for the (fermion) quasi-particles
and the (scalar) plasmon sector is currently under investigation.

\end{document}